\documentclass{article}

\usepackage{PRIMEarxiv}

\usepackage[utf8]{inputenc} % allow utf-8 input
\usepackage[T1]{fontenc}    % use 8-bit T1 fonts
\usepackage{hyperref}       % hyperlinks
\usepackage{url}            % simple URL typesetting
\usepackage{booktabs}       % professional-quality tables
\usepackage{amsfonts}       % blackboard math symbols
\usepackage{nicefrac}       % compact symbols for 1/2, etc.
\usepackage{microtype}      % microtypography
\usepackage{lipsum}
\usepackage{fancyhdr}       % header
\usepackage{graphicx}       % graphics
\graphicspath{{media/}}     % organize your images and other figures under media/ folder

\title{Assessing wind energy's potential for Kentucky
\thanks{\textit{\underline{Citation}}: 
\textbf{Syed A.S. et. al. Assessing wind energy's potential for Kentucky. pp. 1-10}} 
}

\author{
  Abbas Shah Syed  \\
   Department of Electronic Engineering\\
  Mehran University of Engineering and Technology\\
  Jamshoro, 76062, Pakistan\\
  \texttt{zaigham.shah@faculty.muet.edu.pk} \\
   \And
  Aron Patrick \\
  PPL Corporation \\
  PA, 18101-1179, USA\\
  \texttt{ALPatrick@pplweb.com} \\
   \AND
  Adrian Lauf \\
   Department of computer Science and Engineering \\
  University of Louisville \\
  Louisville, KY, 40208 USA\\
   \texttt{adrian.lauf@louisville.edu} \\
   \And
   Adel Elmaghraby \\
    Department of computer Science and Engineering \\
  University of Louisville \\
  Louisville, KY, 40208 USA\\
   \texttt{adel.elmaghraby@louisville.edu} \\
}

\begin{document}
\maketitle

\begin{abstract}
Recently, there has been a push by countries to diversify their energy mix considering various factors. In this regard, there have been several studies conducted to assess the potential for using sources such as wind and solar to generate supplemental energy to the already present energy generation setup. In this regard, this study explores the potential of wind for the Commonwealth of Kentucky. To perform this study, wind data was sourced for eight locations across Kentucky from the publicly accessible wind speed information present at Weatherunderground for the years 2020-2021. An analysis was performed concerning the seasonal, monthly and hourly variation in the wind speed so as to identify the expected times of sufficient wind energy generation. Moreover, a comparison of the collected data was performed with data from a home-based weather station and a deployed wind turbine as well to validate the variation pattern of the publicly sourced data. Finally, in order to investigate the variation patterns of wind and solar energy sources, a comparative analysis was also performed using data from a solar power generation plant in Kentucky.  It was observed that a seasonal and monthly complemetarity was observed between the wind and solar. However, when considering daily patterns, the wind was found to follow solar generation with an offset. While further research is required, this analysis indicates that it is possible to deploy wind energy power generation projects in the Commonwealth of Kentucky. The seasonal complementary behavior of wind and solar can be used along with battery storage in conjunction with natural gas to provide a diversified electricity generation portfolio. 
\end{abstract}

% keywords can be removed

\section{Introduction}

The power requirements of countries around the world are increasing with the growth in human population. The population of the world is expected to increase to 8.5 Billion people by 2030, and will result in an increase in world energy demand by 21\% as predicted by the International Energy Agency \cite{CAPP}. Moreover, the electricity consumption for North America is projected to increase from 4,222 TWh in 2020 to 5,687 TWh in 2050 \cite{electricityforecastconsumption}. Such increases in electric power requirements have been observed in developing countries as well. Therefore, countries around the world have become interested in diversifying their energy mix to cater to growing energy needs. Taking this in to consideration, the aim of the current study are as follows.

\begin{itemize}
    \item Analyze wind data and explore variation patterns to provide insight in to the suitability of wind power generation in Kentucky. 
    \item Analyze the variation pattern of solar energy and compare with wind to assess complemetarity. 
	% \item Identify suitable locations for wind deployment
\end{itemize}

Solar and wind generation systems for electricity can work as part of the grid or in a standalone arrangement and can be used in both rural as well as urban settings \cite{Khare2016}. It therefore has been the focus of multiple studies around around the world \cite{Carvajal2019,adem2022evaluation,AlGhamdi2022,Neupane2022,mahmud2022geographical,slusarewicz2018assessing,bououbeid2024modelling, bououbeid2024modelling, zhang2023overview}. These studies have ascertained the potential of wind energy in different regions of the world. Neupane  et. al in \cite{Neupane2022} studied solar and wind energy potential in Nepal, Al-Ghamdi et. al \cite{AlGhamdi2022} studied wind energy potential and data for various locations in the Kingdom of Saudi Arabia, Adem Cakmakcu and Huner \cite{adem2022evaluation} assessed the potential of wind power generation at a location in Turkey and \cite{mahmud2022geographical,abido2022seasonal} studied the variability of onshore wind in California and in Texas \cite{slusarewicz2018assessing}. The authors aimed to provide an assessment of the useability of wind as an energy generation source. 

Understanding that there is a spectrum of needs in terms of diversifying energy generation sources, this study highlights the case of wind. We start by providing an introduction to the current state of renewable energy generation, focusing on the wind in Section \ref{sec:renewableenergy}. Section \ref{sec:datacollection} presents the data collection scheme used for gathering wind data from different locations in Kentucky. An anlysis of wind speeds at the different locations is presented in Section \ref{sec:windspeedsinkentucky}. Section \ref{sec:solar} discusses the scenario of complementary use of wind and solar energy to achieve a reliable over-the-year energy mix. Finally, the study is concluded in Section \ref{sec:conclusion}.
\section{Energy from Wind}{\label{sec:renewableenergy}}

According to the International Energy Agency, global power generation through renewables was 30.2\% in 2023 \cite{iea_renewables}. In the US, renewable energy generation is expected to more than double from 21\% in 2021 to 44\% in 2050 making it the fastest growing energy source \cite{annualenergyoutlook2022}. This is not surprising given that renewable energy represents the majority of new power added to the generation capacity around the world \cite{10yearsprogress} and is expected to grow by 17\% in 2024 \cite{deloitte2024} .
%Considering the US, the integration of renewable energy in to the North American %Power System can achieve significant carbon reduction (80\%) by the year 2050 \cite{narenewableenergy} thereby making exploration of renewable energy generation and integration a powerful tool to achieve climate change goals in regard to reduction of carbon reduction targets. Moreover, it is envisioned that cooperation between and within countries could produce significant economic benefits as well. 
The integration of renewable energy in to the North American power system can produce significant economic benefits as well.

\begin{figure}[!htb]
	\centering
	\includegraphics[trim=0 0 0 0, scale = 1.2]
 {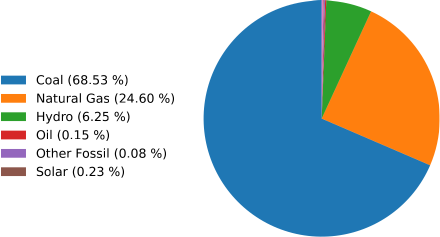}
	\caption{Power Composition for Kentucky.}
	\label{fig:powercompositionky}
\end{figure}

Wind energy has been touted as an important renewable energy source and is expected to be the fastest growing renewable energy source from 2020 through 2024, being the reason for two third of the growth in the renewable energy sector during this time \cite{useiastatistics}. As per data of the US Energy Information Agency (EIA) \cite{wherewindpower}, while the US is responsible for generating 21\% of the worlds wind energy generation, the utilization of wind energy for power generation by Kentucky is well below other states with similar wind characteristics. The wind speed in Kentucky seems similar to most of the east coast and west coast, where as it is higher in the central region of the US. Even with this, utilization potential of wind for power generation in each area is high and most of the benefit is being taken from this resource. Moreover, according to the Office of Energy and Renewable energy, Kentucky currently only produces 0.23\% of its total electricity generation from solar and none from wind \cite{windexchange} with the majority being generated by coal followed by natural gas and hydropower as illustrated in Figure \ref{fig:powercompositionky}\cite{windexchange}.

In fact, according to the US wind turbine database \cite{windturbinedatabase}, there are no wind turbines located in Kentucky which indicates to the underutilization of wind as a power generation source in the state. Therefore, there is a need to explore the potential of using wind as a complementary or supplementary energy source. With government incentives \cite{windincentives} recently being passed, wind-based energy generation can be a potentially viable power generation source.

\section{Data Collection}{\label{sec:datacollection}}

\begin{figure}[!htb]
	\centering
	\includegraphics[trim=0 0 0 0]%, scale = 0.99]
 {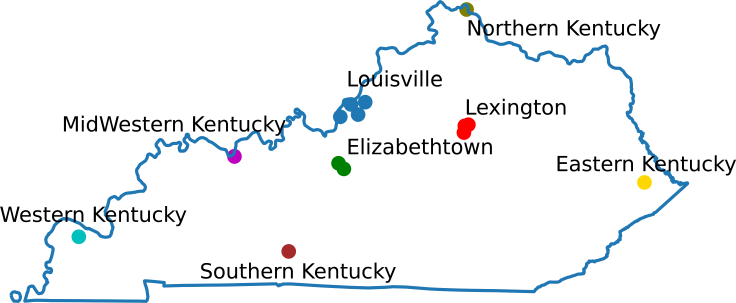}
	\caption{Locations for Wind Stations.}
	\label{fig:stationmap}
\end{figure}

With the aim to analyze the potential for wind power generation throughout Kentucky, data collection was performed in a variety of locations while dividing Kentucky in to different geological areas. Considering data availability constraints, eight locations were chosen, which are the metro areas of Louisville and Lexington, Elizabethtown, Northern Kentucky (NKY), Eastern Kentucky (EKY), Western Kentucky (WKY), Mid-Western Kentucky (MWKY) and Southern Kentucky (SKY). Weather data was scraped from the publicly available weather station website Weatherunderground \cite{weatherunderground}. Weatherunground is a commercial service providing weather information in real-time. The data is sourced from hobbyists as well as commercial weather stations. The station locations from which data was sourced are shown in Figure \ref{fig:stationmap}. 

In order to improve the fidelity of data, multiple stations were considered for the large populated areas of Louisville (four), Lexington (three) and Elizabethtown (two) while for the other locations data from only one station was collected. The data for the stations was updated regularly every five minutes. Also, in order to ensure that a uniform comparison is made, data was collected for the years 2020 and 2021 from beginning of January to the end of December. Moreover, it should be noted that all wind speeds were measured in m/s.

\section{Wind Speeds in Kentucky} \label{sec:windspeedsinkentucky}

In this section, the data collected from Weatherunderground has been analyzed and a discussion is presented on the analysis. 

\subsection{Wind Speeds by Season}
\begin{figure}[!htb]
	\centering
	\includegraphics[trim=100 10 30 10, width = 0.7\textwidth]{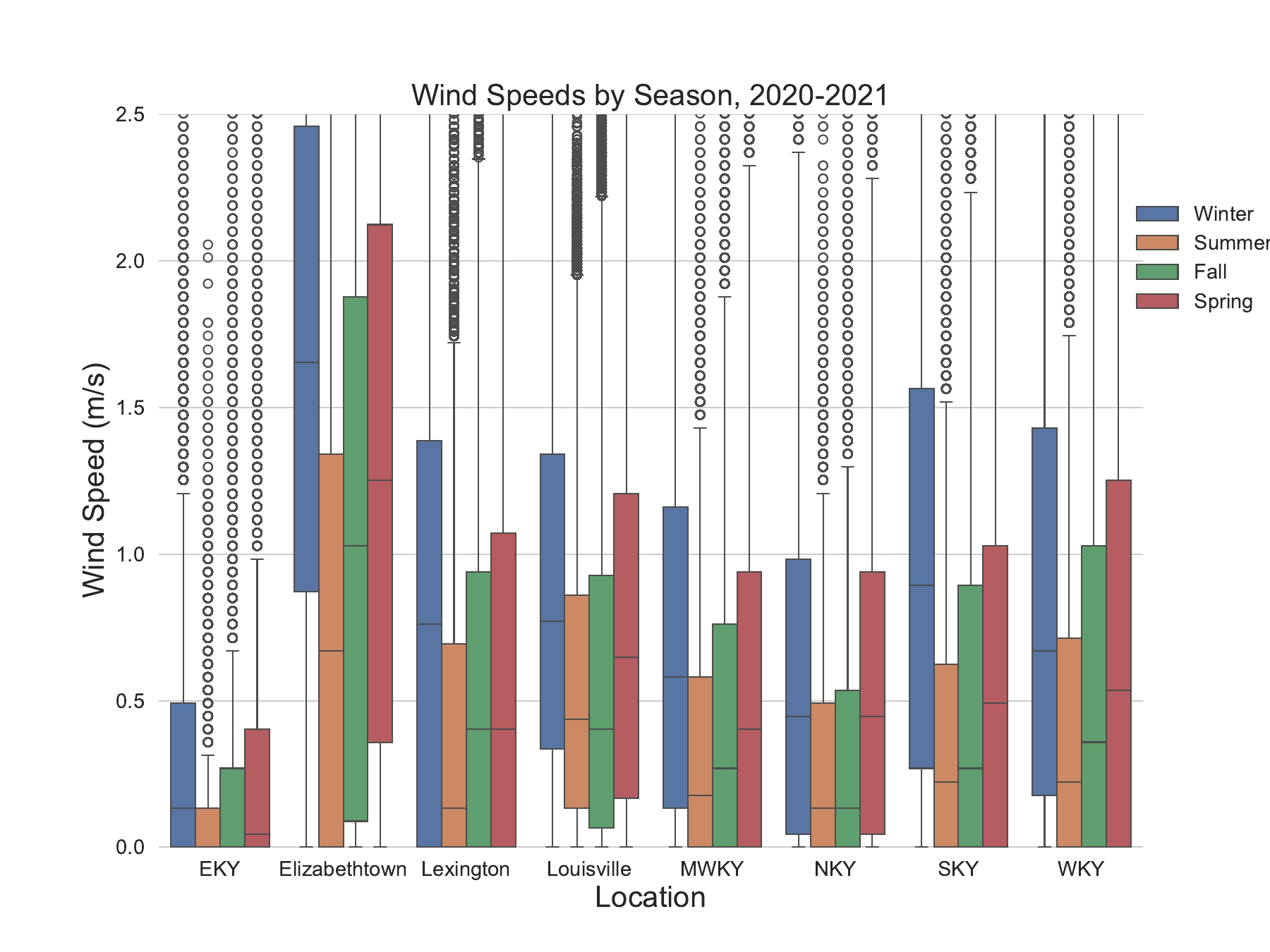}
	\caption{Kentucky Wind Speeds by Season.}
	\label{fig:windspeedboxplots}
\end{figure}

An important aspect to explore when considering the use of wind as a potential source of energy is the variation with respect to season. This is important as energy requirements vary with season. To provide an insight in to the trend of the seasonal variation of wind, Figure \ref{fig:windspeedboxplots} shows the box plots for the wind speed by season for the eight stations. It can be observed from the figure that the highest speeds are observed in winter, followed by spring and fall with the lowest wind speeds observed in summer for all locations considered. 

It can also be observed that from the data collected, the locations at Elizabethtown, Lexington and Louisville and Western Kentucky (WKY) demonstrated the highest wind speeds among the stations whereas the location considered in Eastern Kentuckly (EKY) was the least windiest. 

\subsection{Wind Speeds by Month}

\begin{figure}[!htb]
	\centering
	\includegraphics[trim=150 0 120 0, width = 0.7\textwidth]{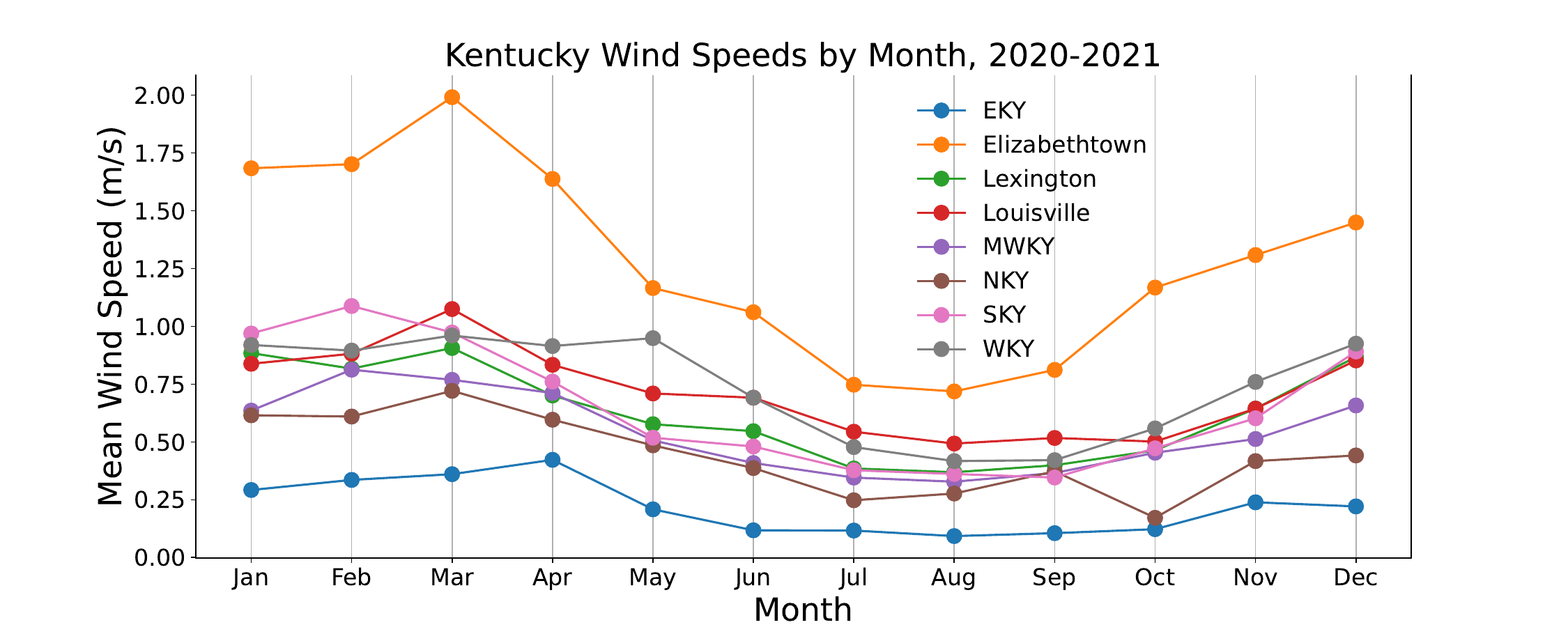}
	\caption{Kentucky Wind Speeds by Month.}
	\label{fig:kywindspeedbymonth}
\end{figure}

 In order to compare the wind speed for all stations on a monthly basis, the plots for the mean wind speed for each month for all stations are shown in Figure \ref{fig:kywindspeedbymonth}. It can be observed that Elizabethtown is the windiest station as observed by month as well. The least windy is Eastern Kentucky. Moreover, it can be observed from the figure that the wind speed is sufficiently high for the months of January – April and November, December. However, for the other months, a below-par speed is observed.

\subsection{Wind Speeds by Day}
In order to understand the daily variation of wind speed, for each location, the average wind speed was computed on a daily basis. From this, Table \ref{tab:windiestday} lists the dates on which the wind was the highest on average for the combined two years considered. It was found that five of the eight locations from which data was analyzed had the windiest day in the spring (end of March, April and beginning of May) whereas three had their windiest day in the winter (end of December, January and beginning of February).

\begin{table}[!htb]
\caption{Windiest Day (on average) for all locations.}
	\label{tab:windiestday}
	\centering
\begin{tabular}{lcc}
\toprule
\textbf{Station/Location} & \multicolumn{1}{l}{\textbf{Date-Month}} & \multicolumn{1}{l}{\textbf{Average Wind Speed (m/s)}} \\ \midrule
\textbf{EKY}              & 1-April                                 & 0.91                                                  \\
\textbf{Elizabethtown}    & 28-March                                & 3.1                                                   \\
\textbf{Lexington}        & 9-May                                   & 1.63                                                  \\
\textbf{Louisville}       & 25-December                             & 1.86                                                  \\
\textbf{MWKY}             & 31-March                                & 1.65                                                  \\
\textbf{NKY}              & 5-February                              & 1.76                                                  \\
\textbf{SKY}              & 31-January                              & 2.49                                                  \\
\textbf{WKY}              & 10-May                                  & 2.81                             \\ \bottomrule
\end{tabular}
\end{table}

\subsection{Wind Speeds by Hour}
\begin{figure}[!htb]
	\centering
	\includegraphics[trim=150 0 120 0, width = 0.7\textwidth]{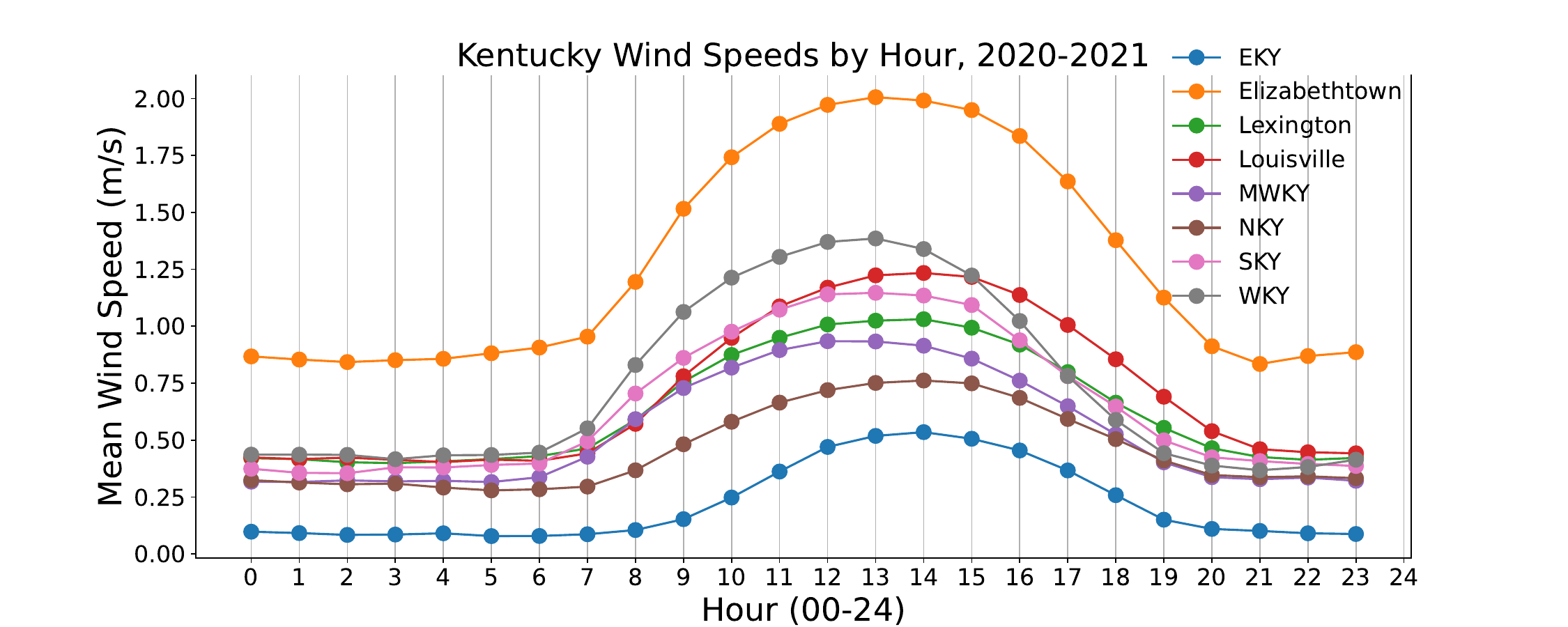}
	\caption{Kentucky Wind Speeds by Hour.}
	\label{fig:hourlywind}
\end{figure}

In order to gauge the trend in the wind speed variation, Figure \ref{fig:hourlywind} illustrates the hourly wind speeds for the locations considered. It can be observed that the highest wind speeds are observed between the hours of 10:00 and 18:00. Such a trend was observed for all the eight station locations considered in this work.

%Figure \ref{fig:monthlywind} shows the monthly wind speed for the top five stations, 

Inorder to better understand the wind speed, Table \ref{tab:windiesthour} lists the windiest hour on average for all considered locations. It was found that the windiest hour for all stations was found to be in the afternoon. Out of them, four stations had 14:00 hours as the windiest hour, three had the most wind on average at 13:00 hours with one station/location having the most wind on average at noon.

\begin{table}[!htb]
\caption{Windiest Hour (on average) for all locations.}
	\label{tab:windiesthour}
	\centering
\begin{tabular}{lcc}

\toprule
\textbf{Station/Location} & \textbf{Hour of the Day (00-24)} & \textbf{Average Wind Speed (m/s)} \\ \midrule
\textbf{EKY}              & 14:00                            & 0.54                              \\
\textbf{Elizabethtown}    & 13:00                            & 2.01                              \\
\textbf{Lexington}        & 14:00                            & 1.03                              \\
\textbf{Louisville}       & 14:00                            & 1.23                              \\
\textbf{MWKY}             & 12:00                            & 0.93                              \\
\textbf{NKY}              & 14:00                            & 0.76                              \\
\textbf{SKY}              & 13:00                            & 1.15                              \\
\textbf{WKY}              & 13:00                            & 0.54                           \\ \bottomrule
\end{tabular}
\end{table}

%\subsection{Wind Speeds at 5-minute intervals}

%In order to better appreciate the variation in the windspeeds, the wind speed at a resolution of five minute measurements has been shown in Figure \ref{fig:windspeed5min} for the day with the highest wind speed for the considered locations. It should be noted that the days are different for each location. 

%It can be observed from the figure that the wind speed does vary quite a lot even with respect to five minute intervals. This behavior is an important consideration when considering the deployment of wind power systems. 

%\begin{figure}[!htb]
	%\centering
	%\includegraphics[trim=150 0 50 0, width = 0.8\textwidth]{Figures/%windspeed5min.pdf}
	%\caption{Wind Minute (5 minute resolution) for top five locations for day of %maximum average capacity factor.}
	%\label{fig:windspeed5min}
%\end{figure}

\subsection{Wind Data Comparison}

In this section, the publicly collected wind data from Weatherunderground has been compared with two different data sources. The first is a home-based weather station in a suburb in Louisville, Kentucky and the other is a Wind Turbine in Louisville, which also provided wind measurements. This comparison is presented to highlight the patterns of observed wind speed variation.

\subsubsection{Home-based Weather Station Data}

\begin{figure}[!htb]
	\centering
	\includegraphics[trim=150 0 120 0, width = 0.7\textwidth]{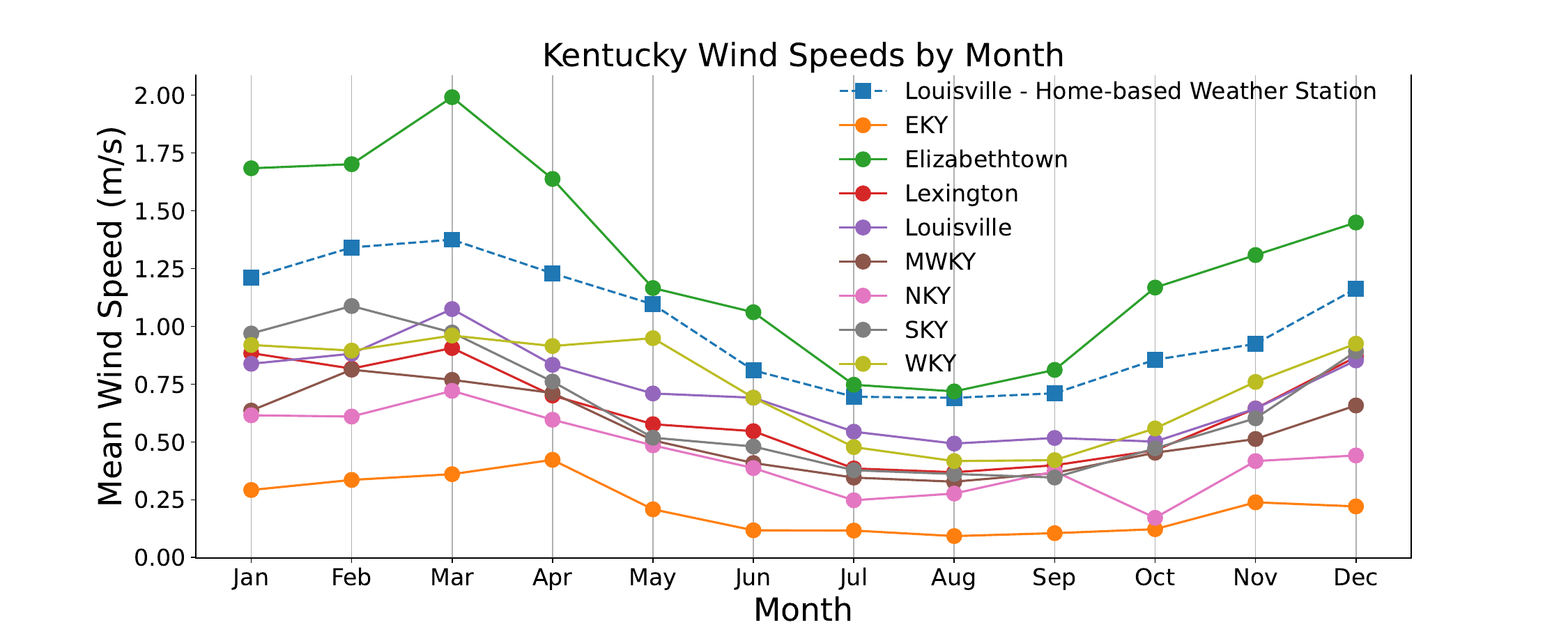}
	\caption{Wind Speeds by Month - Comparison with Home-based Weather Station}
	\label{fig:drlaufmonth}
\end{figure}

\begin{figure}[!htb]
	\centering
	\includegraphics[trim=150 0 120 0, width = 0.7\textwidth]{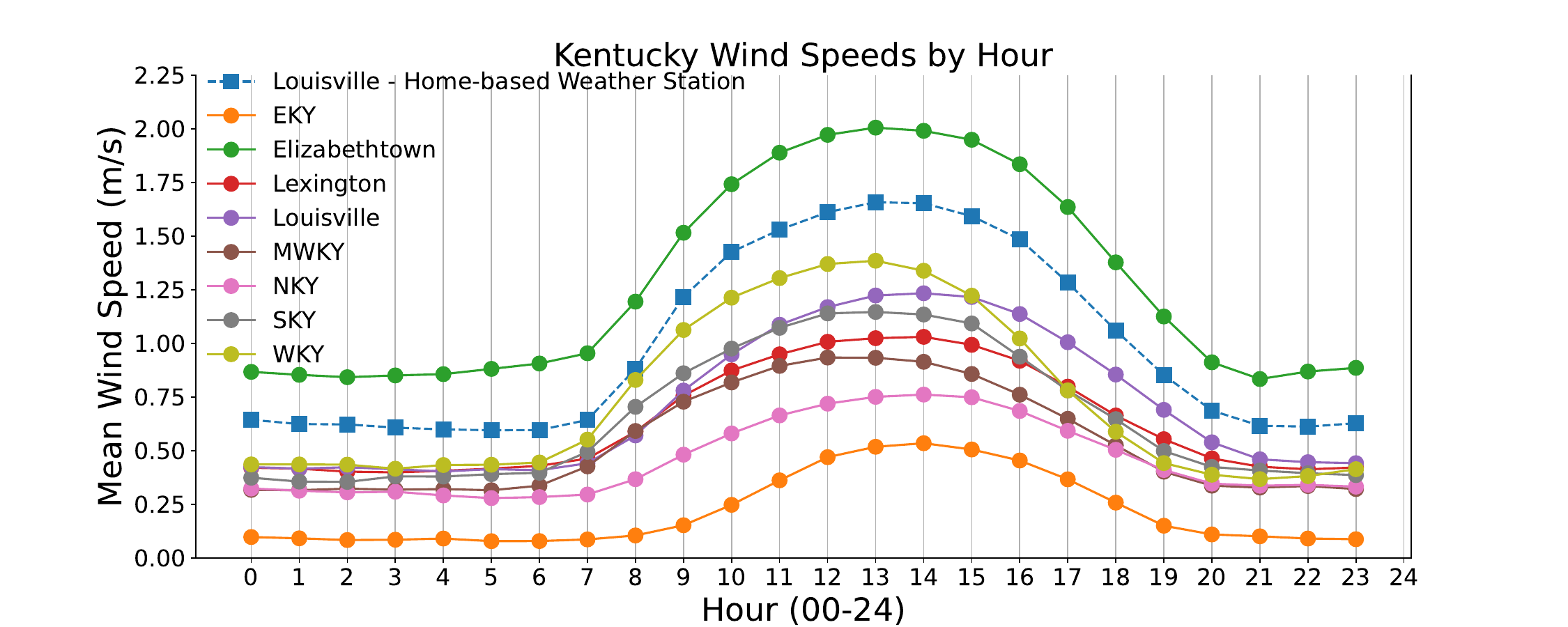}
	\caption{Wind Speeds by Month - Comparison with Home-based Weather Station}
	\label{fig:drlaufhour}
\end{figure}

In order to compare the wind pattern from the data analyzed, data from a home-based weather station was acquired for the time period between 1st July 2021 to 15 March 2023. This was then resampled for by month and hour. The plot for monthly wind speeds has been illustrated in Figure \ref{fig:drlaufmonth}. Moreover, the plot for hourly wind speeds has been illustrated in Figure \ref{fig:drlaufhour}. 

From Figure \ref{fig:drlaufmonth} and Figure \ref{fig:drlaufhour} it can be observed that the pattern for wind speed variations are very similar for the collected data and the data acquired from the home-based weather station. Moreover, the fact that the data from the home-based weather station contains data outside the collected data's time frame (1 January 2020 to 31 December 2021) attests to the dependability of the wind as an energy-producing resource. 

\subsubsection{Wind Turbine Data}

\begin{figure}[!htb]
	\centering
	\includegraphics[trim=150 0 120 0, width = 0.7\textwidth]{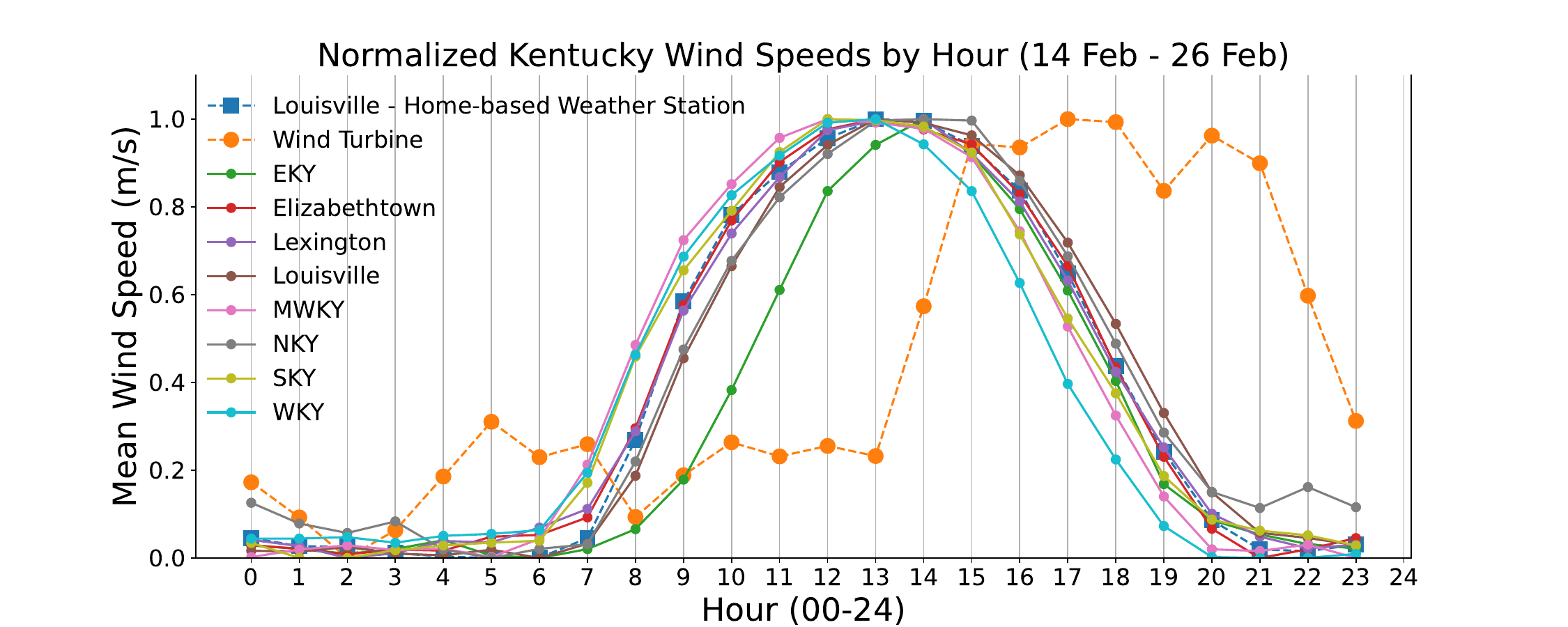}
	\caption{Wind Speeds by Month - Comparison Wind Turbine Measurements}
	\label{fig:winddrlaufhourly}
\end{figure}

In order to further compare the wind pattern from the data analyzed, data from a wind turbine in Louisville was acquired. The data contains wind speed measurements for twelve days between 14 February 2024 to 26 February 2024. In order to understand the pattern better, only the collected data from Weatherunderground between the period of 14 February and 26 February but between the years 2020 and 2021 has been utilized. The plot for hourly wind speeds has been illustrated in Figure \ref{fig:winddrlaufhourly}. It should be noted that the wind speeds have been normalized to 0 and 1 to attain a fair comparison. 

It can be observed from Figure \ref{fig:winddrlaufhourly} that there is a slight offset in terms of the pick-up of wind during the day in the wind turbine data as compared to the data collected from both from Weatherunderground as well as the data acquired from the home-based weather station.

\section{Solar Energy}\label{sec:solar}

One disadvantage that energy sources like wind have is fluctuation in their availability. However, power generation by wind can be complemented with solar power generation to provide a more dependable source of energy. In order to provide an assessment of this possibility, solar power generation data from the LGE E.W. Brown Solar Facility \cite{livesolardata} is used. The data was acquired for the period of 1 January 2020 to 31 December 2021. 

\subsection{Solar by Season}
\begin{figure}[!htb]
	\centering
	\includegraphics[trim=0 0 0 0, width = 0.65\textwidth]{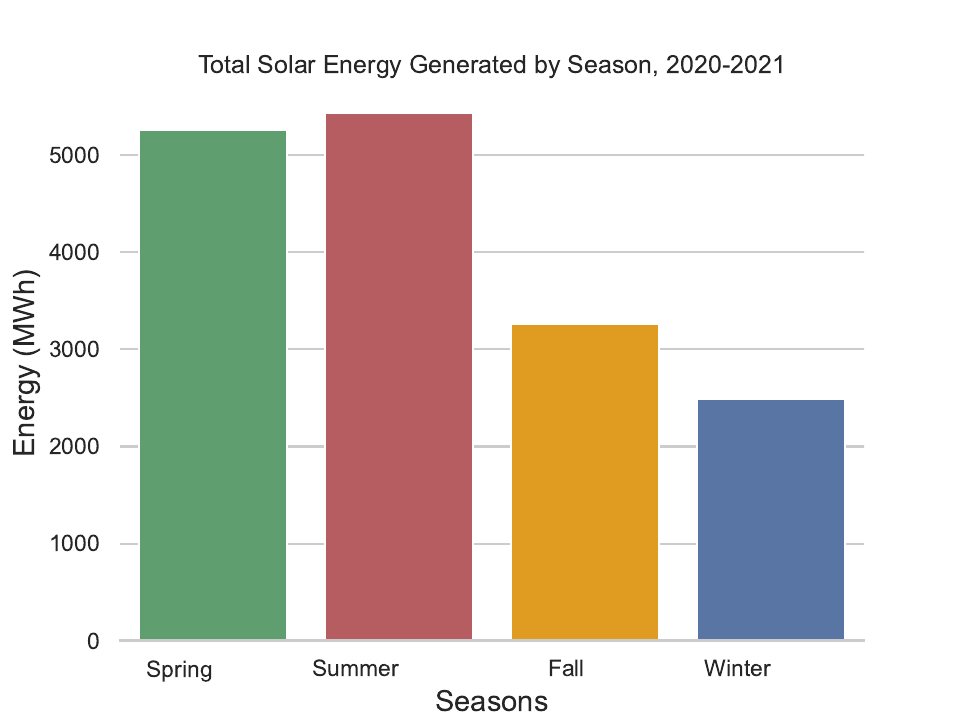}
	\caption{LGE Solar Power Plant Energy Production between 2020-2021.}
	\label{fig:solarseasonal}
\end{figure}

The seasonal total energy generation by the facility is shown in  Figure \ref{fig:solarseasonal}. It can be observed that the most energy is generated around summer time with the energy around the winter decreasing drastically. This is opposite behavior when compared to the prevalence of wind (Figure \ref{fig:windspeedboxplots}) in terms of speed, the peak season of wind speed is observed during winter. Moreover, it can also be observed from the figure that the amount of energy generated for spring and summer is very similar. 

% The total annual energy generation is close to 16,500 MWh for each year for the two years considered. 

\subsection{Solar-Wind by Month}

\begin{figure}[!htb]
	\centering
	\includegraphics[trim=130 0 100 0, width = 0.71\textwidth]{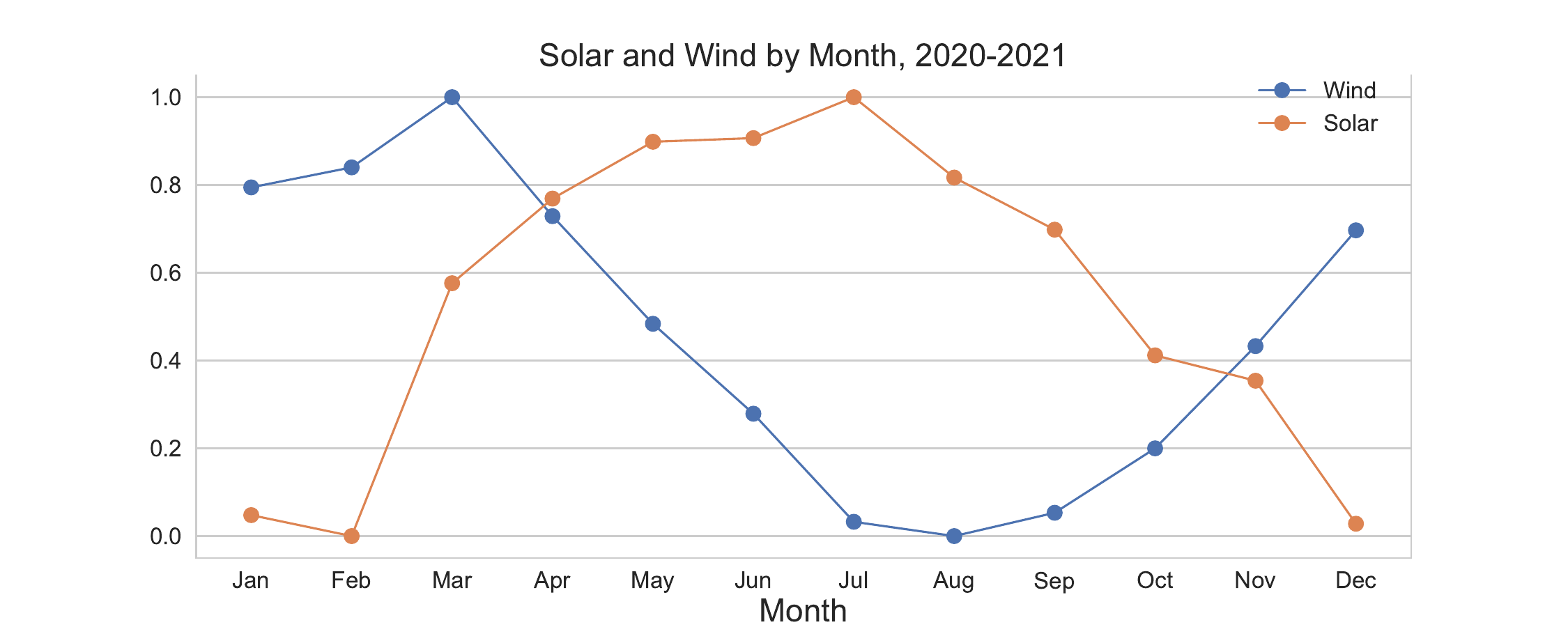}
	\caption{Solar and Wind Variation by Month between 2020-2021.}
	\label{fig:solarwindmonth}
\end{figure}

In order to further explore the relation between using Solar and Wind as energy generation sources, wind speed data collected from Weatherunderground across all stations together has been averaged with respect to month and normalized. A similar aggregation for solar power is also performed, the aim being to observe the pattern of variation of the two sources together. The plots are shown in Figure \ref{fig:solarwindmonth}. It can be observed that the peak of wind occurs in the month of March whereas the peak of solar occurs in the month of July and both sources are complementary at this time. There is an overlap in the spring and the fall months when the transition between the seasons takes place.

\subsection{Solar-Wind by Hour}

\begin{figure}[!htb]
	\centering
	\includegraphics[trim=130 0 100 0, width = 0.71\textwidth]{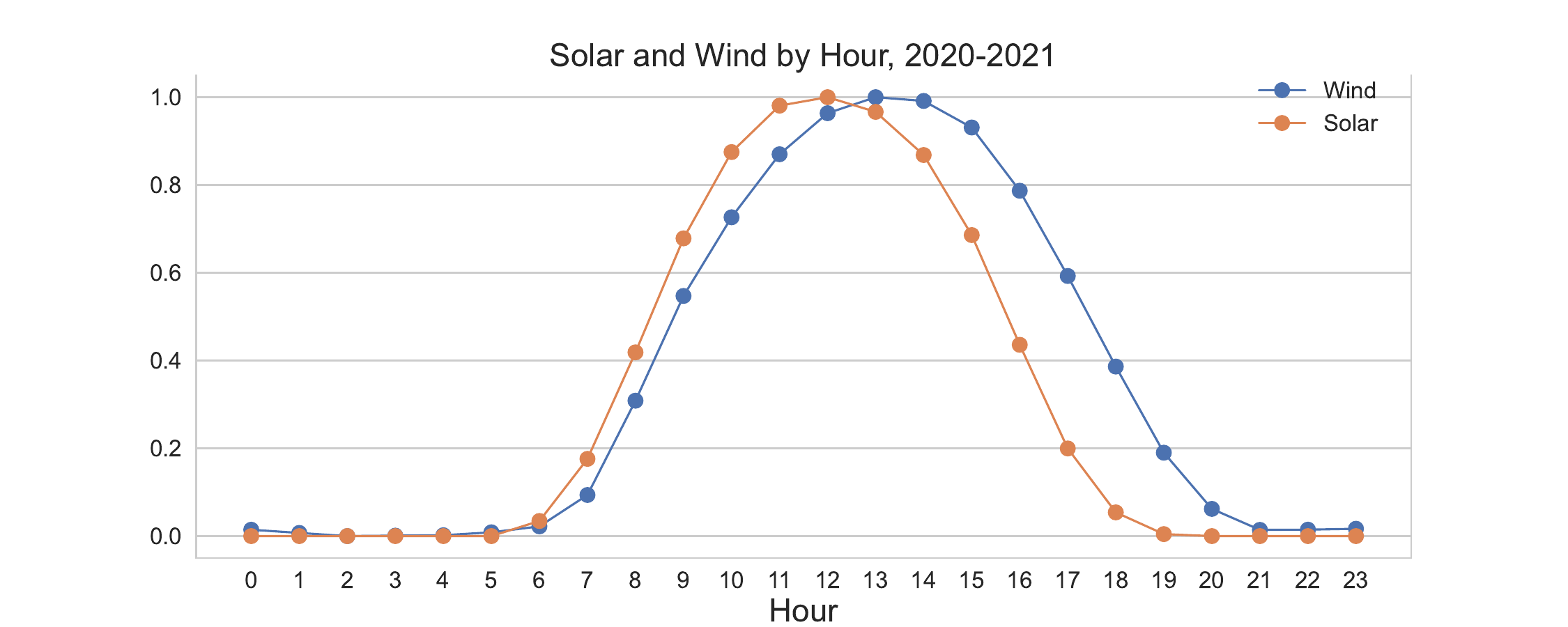}
	\caption{Solar and Wind Variation by Month between 2020-2021.}
	\label{fig:solarwindhour}
\end{figure}

For a more granular analysis, the wind and solar data was resampled for every hour and then normalized. This is carried out to investigate the comparative patterns during the day. The plots are illustrated in Figure \ref{fig:solarwindhour}. It can be observed that the wind and solar are offset by a few hours. A high of solar energy is followed by high wind speed. This is due to convection, where warmth from the sun creates an upward movement of warm air. Dense or cooler air from the surroundings then moves into this low-pressure area to fill in the void and therefore gives rise to an increase in wind speed.

\section{Conclusion}\label{sec:conclusion}

The current work provides an exploratory study into the wind patterns for the Commonwealth of Kentucky. To perform the analysis, data was sourced from the publicly accessible Weatherunderground portal for eight different locations across Kentucky, covering the three urban areas of Louisville, Lexington, and Elizabethtown as well as the regions of Eastern Kentucky, Southern Kentucky, Western Kentucky and MidWestern Kentucky. An Analysis was then performed with different granularities, including seasonal, monthly, and hourly to investigate the pattern of variation at the different locations considered. It was found that the wind speeds were the highest in the winter, followed by spring while they were the least in summer. When looking at the month, all locations had wind speeds being the highest either in the winter or spring months. Considering wind speeds on an hourly basis, it was observed that the wind speed picked up around morning time and decreased in the evening.  Moreover, the pattern of the collected data was compared with data acquired from a home-based weather station as well as a deployed wind turbine. It was observed that the pattern of wind variation was very similar from all data sources.

Furthermore, in order to assess the complementarity of wind with solar, data was acquired from a solar power plant in Kentucky. The energy generated by the solar power plant was taken as an indicator of the solar power availability. A seasonal and month-wise comparison revealed that wind and solar are complementary of each other in terms of each having peaks and troughs at opposite times of the year. Finally, the daily variations of wind and solar energy were also analyzed to ascertain the daily variations between them. It was observed that the wind was produced at an offset of the solar pattern.

Based on the study here from a variety of locations across Kentucky, our analysis of wind data across Kentucky indicates that wind is a potential source of low-carbon electricity generation to seasonally complement solar and integrate with natural gas combined cycle for a more diversified and resilient electricity portfolio. 

\section*{Acknowledgments}
This work was supported in part by a grant from LG\&E.

%Bibliography
\bibliographystyle{unsrt}  
\bibliography{windpaper}

\end{document}